\documentclass[a4paper,12pt]{article}
\pdfoutput=1 

\usepackage{jheppub} 

\usepackage[T1]{fontenc}
\usepackage[english]{babel}
\usepackage{hyperref}
\usepackage{ifpdf}
\usepackage{subfigure}
\usepackage{amssymb}
\usepackage{amsfonts}
\usepackage{epsf}
\usepackage{rotating}
\usepackage{graphicx}
\usepackage{amsmath}
\usepackage{fancyhdr}
\usepackage{babel}
\usepackage{graphics}
\usepackage{pstricks}
\usepackage{color}
\usepackage{multirow}

\newcommand{\nn}{\nonumber}
\newcommand{\lsim}{\mathrel{\mathop{\kern 0pt \rlap
  {\raise.2ex\hbox{$<$}}}
  \lower.9ex\hbox{\kern-.190em $\sim$}}}
\newcommand{\gsim}{\mathrel{\mathop{\kern 0pt \rlap
  {\raise.2ex\hbox{$>$}}}
  \lower.9ex\hbox{\kern-.190em $\sim$}}}

\newcommand{\be}{\begin{equation}}
\newcommand{\ee}{\end{equation}}
\newcommand{\bea}{\begin{eqnarray}}
\newcommand{\eea}{\end{eqnarray}}

\def\etmiss{\not\!\!{E_T}}

\title{\boldmath Displaced lepton flavour violating signatures of right-handed sneutrinos in $U(1)'$ supersymmetric models }

\author[a]{Priyotosh Bandyopadhyay}

\affiliation[a]{Dipartimento di Matematica e Fisica "Ennio De Giorgi", Universit\`a del Salento and INFN-Lecce, Via Arnesano, 73100 Lecce, Italy\\
and\\
Indian Institute of Technology Hyderabad, Kandi,  Sangareddy-502287, Telengana, India}
\emailAdd{priyotosh.bandyopadhyay@le.infn.it, bpriyo@iith.ac.in}

\abstract{We consider a $U(1)'$ extended supersymmetric model with a right-handed 
neutrino superfield which can generate light neutrino mass by Type I seesaw mechanism.
The lighter superpartner of the right-handed neutrino could be the scalar  dark matter.
These right-handed sneutrinos can come from the decay of $\tilde{Z'}$, superpartner of the
extra gauge boson $Z'$. Left-right handed sneutrino mixings affect their decays further, giving
rise to displaced "lepton flavour violating" signatures.  A wino-like chargino NLSP (next to lightest supersymmetric particle)  creates even more interesting decay topology. We investigate such displaced multi-leptonic final states with "lepton flavour violation" from the  supersymmetric cascade decays of third generation squarks at the LHC.}


\begin{document}

\maketitle
\flushbottom
\section{Introduction}

The recent discovery of a Higgs boson around $125$ GeV has established the role of 
at least one scalar in the electro-weak symmetry breaking \cite{LHCH}. However, experimental evidences for tiny neutrino masses and indirect evidence of dark matter candidate, and a theoretical requirement for naturalness of the electro-weak scale need a theory beyond the Standard Model (SM). A supersymmetric seesaw model \cite{Mohapatra05} with an additional $U(1)'$  \cite{Langacker08} gauge symmetry can address these issues consistently. 

In supersymmetric theories with R-parity, the lightest supersymmetric particle (LSP) is stable and thus 
a neutral LSP, typically a linear combination of neutral gauginos and Higgsinos,  
becomes a good thermal dark matter (DM) candidate if supersymmetry (SUSY) is broken around the TeV scale \cite{Jungman95}. Similarly, $U(1)'$ breaking can also be induced via quantum effects at ${\cal O}$(TeV) \cite{Khalil07} which then generates the seesaw scale around TeV, testable  at LHC. The additional $U(1)'$, arising from any grand unified gauge group, requires the presence of right-handed neutrinos (RHNs) for the anomaly-free condition.

A right-handed sneutrino (RHsN) can be  a good  scalar dark matter candidate
with correct thermal relic if the $U(1)'$ gaugino $\tilde Z'$, the superpartner of 
the $U(1)'$ gauge boson $Z'$, is relatively light \cite{Bandyo11}. 
The scenario gets very interesting when one of the RHsNs is LSP with a chargino next to LSP (NLPS) and the other one is next to next LSP (NNLSP). The RHsNs being SM gauge singlets can evade the recent bounds on NLSP-LSP \cite{chgn1}.  The same reason leads to the unusual decays of lighter chargino and RHsNs which are  discussed in this article. We consider the $U(1)'$ extended supersymmetric scenario with an right handed sneutrino-superfield. The characteristics of such decays are rather general and though we consider $U(1)_{\chi}$ extended supersymmetric scenario for the analysis, such observations are  feasible in other $U(1)'$ extensions \cite{Langacker08}.

In this article we show the possible  LHC signatures of the displaced "lepton flavour violating" decays
of RHsN caused by left-right handed sneutrino mixings. Being SM gauge singlet and only charged 
under $U(1)_{\chi}$, such RHsNs will mostly come from $\tilde Z'$ decay. The pair production cross-section of such $\tilde Z'$ is not so promising \cite{Bandyo11} but its production from SUSY cascade could still be encouraging, in particular from the lighter third generation squark ($\tilde{t}_1, \tilde{b}_1$) decays.  When  the heavier RHsN ($\tilde{N}_2$) is NNLSP  and the lighter one ($\tilde{N}_1$) is  LSP along with wino-like  chargino  NLSP, the situation can lead to multi-leptonic final states along with "lepton flavour violating" displaced charged track. 

In section~\ref{model} we briefly discuss the model and in section~\ref{mix} we describe the mixings between left and right handed sneutrinos.  We discuss the decay phenomenology of $\tilde{Z}'$ in section~\ref{pheno} and perform a collider study for some benchmark points in section~\ref{cold}. Finally, we conclude in section~\ref{concl}.

\section{The Model}\label{model}

We consider  the $U(1)_\chi$ model for our explicit analysis as in \cite{Bandyo11, Bandyo14,Chun:2015oya} and the particle content is as follows:
\begin{equation}\label{chargeqn}
\begin{array}{c|ccc|cc|c|cc}
SU(5) & 10_F & \bar{5}_F & 1 (N) & 5_H & \bar{5}_H & 1 (X) & 1
(S_1) & 1 (S_2) \cr \hline 2 \sqrt{10} Q' & -1 & 3 & -5 & 2 & -2 &
0 & 10 & -10 \cr
\end{array}
\end{equation}
where $SU(5)$ representations and $U(1)'$ charges of the SM
fermions ($10_F, \bar{5}_F$),  Higgs bosons ($5_H,
\bar{5}_H$), and additional singlet fields ($N, X, S_{1,2}$) are
shown. Here $N$ denotes the right-handed neutrino, $X$ is an
additional singlet field fit into the 27 representation of $E_6$,
and we introduced more singlets $S_{1,2}$, vector-like under
$U(1)_{\chi}$, to break $U(1)_{\chi}$ and generate the Majorana mass term of
$N$ \cite{Khalil07}. Note that the right-handed neutrinos carry the largest charge under
$U(1)_{\chi}$ and thus the corresponding $Z'$ decays dominantly to
right-handed neutrinos. The $U(1)_{\chi}$  breaking can be generated radiatively in the minimal
Higgs sector  \cite{Khalil07} alternatively the additional singlet field
$X$ is neutral under $U(1)_{\chi}$ so that it can be used to
generate a mass for the $U(1)_{\chi}$ Higgsinos in non-minimal Higgs sectors \cite{NMSSM}.
Here we do not assume any grand unification theory as the origin of our 
model and the grand unification structure is just  used for a convenient guide to a theoretically consistent 
model guaranteeing the anomaly free condition.

The gauge invariant superpotential in the seesaw sector is given by
\begin{equation}\label{wpot}
W_{seesaw} = y_{ij} L_i H_u N_j + {\lambda_{N_i}\over2} S_1 N_i
N_i\;,
\end{equation}
where $L_i$ and $H_u$ denote the lepton, Higgs doublet superfields and $y_{ij}$, $\lambda_{N_i}$ are the dimensionless couplings respectively. Given the vacuum expectation values $<S_{1,2}>$, the $Z'$ mass is generated as $2g'\sum_{i} Q'^2_{S_i}<S^2_{i}>$, which gets $\gsim 2.5$ TeV bound depending on $U(1)'$ types \cite{z'bound}.

\section{Sneutrino mixings and displaced decay }\label{mix}
The $U(1)'$  symmetry is broken via  $S_1$ acquiring  vev and this also 
generates the Majorana mass terms for the right-handed neutrinos (RHNs), i.e., $m_{N_i}=\lambda_{N_i} \langle S_1
\rangle$. The Majorana mass terms along with electro-weak vev of the up-type Higgs doublet $H_u$, induce the masses for the light neutrinos via Type-I seesaw mechanism as shown below:
\begin{equation}\label{numass}
 \widetilde{m}^\nu_{ij} = - y_{ik} y_{jk}{ \langle H_u^0\rangle^2 \over
 m_{N_k} }\,.
\end{equation}
The same vev also instigates the mixing between the right-handed neutrino $N$ and left-handed neutrino $\nu$
and corresponding mixing angle is given by
\begin{equation}\label{numix}
\theta_{N\nu}\simeq \frac{y_{\nu}v_u}{\sqrt{2}m_N},
\end{equation}
which leads to $N$ decay to SM gauge bosons and leptons.
The real and imaginary components of the scalar part of $N$ superfield, i.e., $\tilde{N}=(\tilde{N}_1 + \tilde{N}_2)/\sqrt{2}$, get a mass splitting proportional to the lepton number violating (Majorana) mass term $m_N$\cite{Bandyo11}. The mixings also occur between the real and imaginary components of the left- and right-sneutrinos proportional to the vev $v_u$ as given by Eq.~\ref{snumix} 
\begin{equation}\label{snumix}
\theta_{\tilde{N}_{1,2} \tilde{\nu}_{R,I}}=  \frac{y_{\nu}(m_N \pm X_L)v_u}{\sqrt{2}(m^2_{\tilde{N}_{1,2}}-m^2_{\tilde{\nu}_{R,I}})}; \, X_L=A_L \mp \mu \cot{\beta},
\end{equation}
where $m_{\tilde{\nu}_{R,I}}$ are the masses  for the real and imaginary part of the left-handed sneutrino, $A_L$ is the tri-linear soft term corresponding to the leptonic doublet, $\mu$ is the up and down type Higgs mixing parameter in the superpotential ($\mu H_d . H_u$) and $\tan{\beta}$ is the ratio of their vevs as explained in \cite{Bandyo11}. 

The mixing angles can be naturally small due to small Yukawa coupling, $y_{\nu} \sim 10^{-7}$ which corresponds to 
a light neutrino mass scale of  $\widetilde{m}_\nu=0.05$ eV for $m_N\sim 100$ GeV. The other possibility 
of very small mixing angles, comes from the cancelation among the parameters in the numerator of Eq.~\ref{snumix}. We now see how these mixing angles affect the decays of lighter chargino and sneutrinos. 

 \begin{figure}[h]
\begin{center}
\includegraphics[width=.61\linewidth]{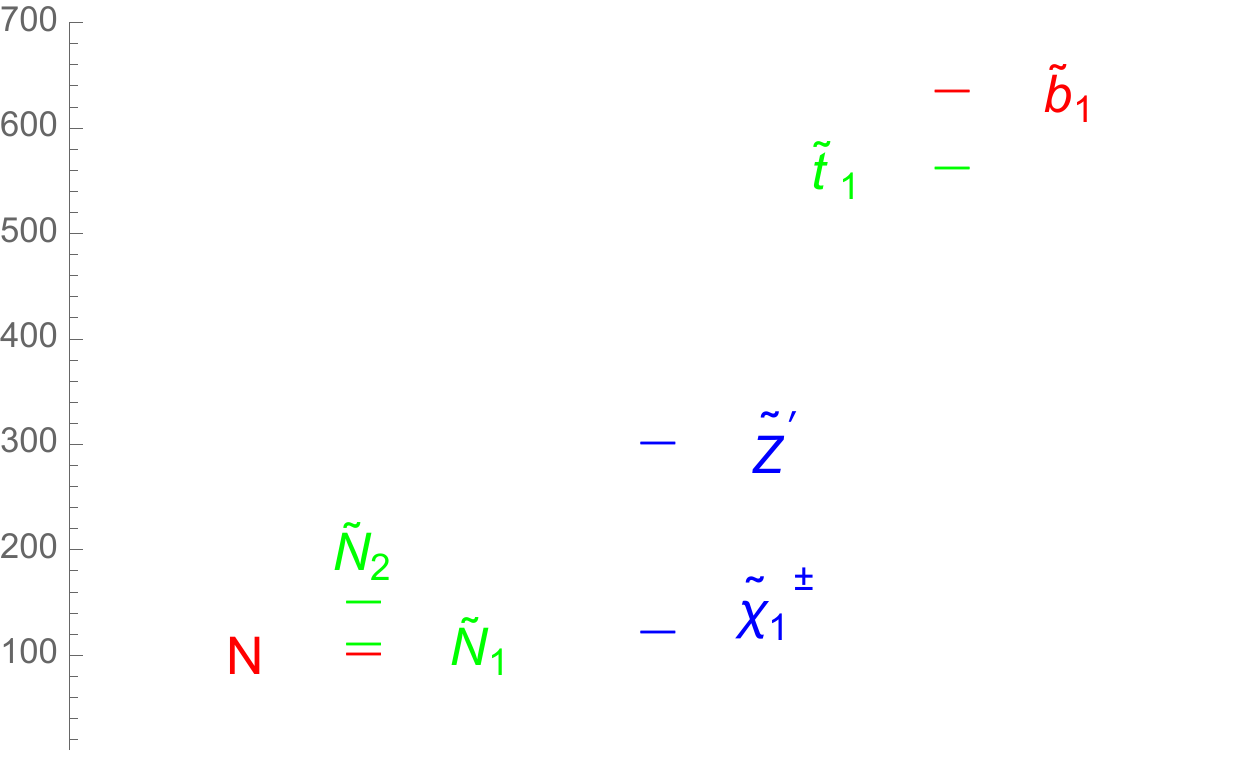}
\caption{A typical mass spectrum for the cascade decay  of $\tilde{Z}'$ via the mixing of left and  right-handed sneutrinos.}\label{massspec}
\end{center}
\end{figure}
Figure~\ref{massspec} shows a typical mass spectrum, where lighter stop and sbottom can decay to 
$\tilde{Z}'$ which further decays to  RHN, RHsNs ($N, \, \tilde{N}_{1,2}$). The lighter sneutrino, $\tilde{N}_1$ is the LSP and a dark matter candidate \cite{Bandyo11}. The wino-like chargino  $\tilde{\chi}^\pm_1$ is the NLSP and the heavier sneutrino $\tilde{N}_2$ stays as NNLSP.  $\tilde{Z}'$ can be produced from the decays of 
$\tilde{t}_{1}$ and $\tilde{b}_{1}$ \cite{Bandyo11, Bandyo14}.

The choice of electro-weak scale mass of $\tilde{Z}'$ is required to  have the correct DM relic in a scenario where the lighter RHsN $\tilde{N}_1$ is DM candidate \cite{Bandyo11}. Both the right-handed superfield and DM are charged under the $U(1)_{\chi}$ as can be seen from Eq.~\ref{chargeqn} and otherwise both are SM gauge singlets. The lightest superpartner of $N$, i.e., $\tilde{N}_1$ is a real scalar field which does not have any $s$-channel annihilation diagram via $Z'$ and $h$ (See couplings in Table1 of \cite{Bandyo11}). These leave no choice other than a $t$-channel annihilation of $\tilde{N}_1 \tilde{N}_1 \to N N$ via  $\tilde{Z}'$    \cite{Bandyo11}. In this process, the decays and inverse decays of the right- handed neutrino, $N$ through the small Yukawa coupling play an important role in keeping the right-handed neutrino in thermal equilibrium and thus controlling the dark matter relic density, which is a distinguishable feature of the thermal history of the right-handed sneutrino dark matter compared with the conventional neutralino LSP dark matter. The $t$-channel annihilation of $\tilde{N}_1 \tilde{N}_1 \to N N$ via  $\tilde{Z}'$ and $N$ decays make the masses of $\tilde{N}_1, \tilde{Z}' \rm{and} \,N$ correlated. This results into very light electro-weak scale $\tilde{Z}'$ masses for a given electro-weak scale  masses $\tilde{N}_1$ and $N$ \cite{Bandyo11}. Thus the mass spectrum and their phenomenology studied in this context is very specific to the choice of the dark matter candidate, in our case which is the lightest right-handed sneutrino $\tilde{N}_1$.
 \begin{figure}
\begin{center}
\includegraphics[width=.51\linewidth]{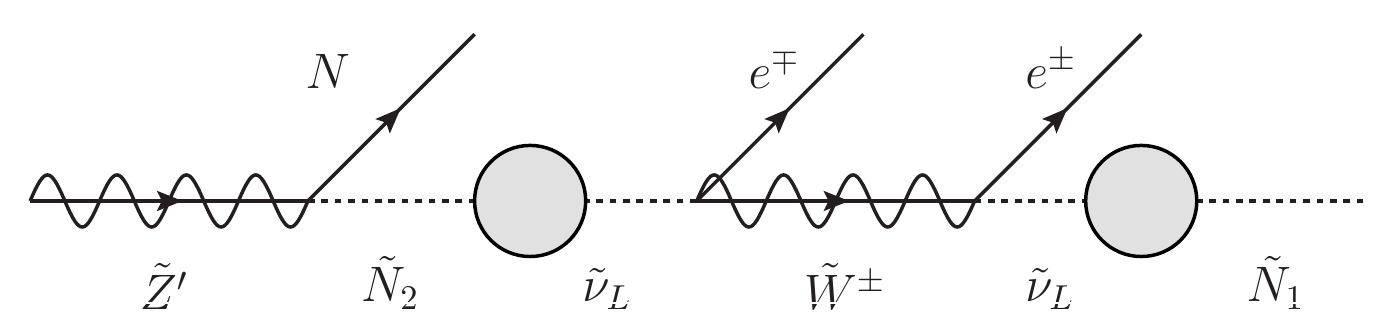}
\caption{Possible cascade decay  of $\tilde{Z}'$ via the mixing of left and  right-handed sneutrinos.}\label{fig:fy}
\end{center}
\end{figure}

The Majorana mass term in Eq.~\ref{wpot} violates the lepton number. However the decay $N \to \ell^\pm W^\mp$ maintains the lepton number. For the current collider analysis we have taken only one $N$ superfield, where $N$ decays into $e^\pm W^\mp$ but not to $N \to \mu^\pm W^\pm$ and $N\to \tau^\pm W^\mp$\footnote{Other two right-handed neutrinos are decoupled and the right-handed sneutrino dark matter annihilates only to the lightest  right-handed neutrino pair in obtaining correct dark matter relic \cite{Bandyo11}.}. We have chosen only one RHN, with the lepton flavour number of 'e'. This preference of one flavour over another can be seen in collider searches as flavour violating decays. From now on the "lepton flavour violation" phrase has been used in their collider search context only.

The effect of kinetic mixing  of $Z$ and $Z'$ is not considered here as their effects are not potential threat to change the mass hierarchy.
This is due to the fact that the right-handed neutrino superfield is singlet under SM gauge groups thus does not couple to $Z$ and only couples to $Z'$. Such kinetic mixing must appear twice in the vacuum polarisation diagram which affects the loop mass proportional to the quadratic power of the mixing angle. Those vacuum polarisation diagrams are further suppressed by the three gauge boson (gaugino) and one RHN (RHsN) propagators. The bounds on the  $Z-Z'$ kinetic mixing and its phenomenology can be found out \cite{kinmx}.

Figure~\ref{fig:fy} depicts the decay of $\tilde{Z}'$ which leads to the scenario we are interested in (shown in Figure~\ref{massspec}). Once a relatively lighter $\tilde{Z}'$ produced, preferably by SUSY cascade decays,  it decays into  a RHN and RHsNs ($\tilde{Z}' \to N \tilde{N}_{1,2}$). $\tilde{N}_2$  then undergoes a "lepton flavour violating" decay to an electron and chargino via the mixing $\theta_{\tilde{N}_2 ,\tilde{\nu}_I}$.  Such chargino undergoes another "lepton flavour violating" two-body decay to $e^\pm \tilde{N}_1$ via  sneutrino mixing $\theta_{\tilde{N}_{1} \tilde{\nu}_{R}}$. 

The decay width of $\tilde{N}_2$ and  $m_{\tilde{\chi}^\pm_1}$ are shown below by Eq.~\ref{dcn2} and Eq.~\ref{chng1} respectively,

\begin{equation}\label{dcn2}
\Gamma(\tilde{N}_2\to e\tilde{\chi}^\pm_1)= \frac{g^2_2}{16\pi}|V_{i1}|^2\theta^2_{\tilde{N}_2,\tilde{\nu}_{I}}m_{\tilde{N}_2}\left(1-\frac{m^2_{\tilde{\chi}^\pm_1}}{m^2_{\tilde{N}_2}}\right)^2,
\end{equation}

\begin{equation}\label{chng1}
\Gamma(\tilde{\chi}^-_1\to e\tilde{N}_1)= \frac{g^2_2}{32\pi}|V_{i1}|^2\theta^2_{\tilde{N}_1,\tilde{\nu}_{R}}m_{\tilde{\chi}^\pm_1}\left(1-\frac{m^2_{\tilde{N}_1}}{m^2_{\tilde{\chi}^\pm_1}}\right)^2
\end{equation}

where $V$ is one of the chargino mixing matrices, $g_2$ is the $SU(2)$ gauge coupling, $m_{\tilde{\chi}^\pm_1}$ is the lighter chargino mass and  $\theta_{\tilde{N}_2 ,\tilde{\nu}_I}$ can be calculated from Eq.~\ref{snumix}. We can see from  Eq.~\ref{snumix} that for some parameter space, either $\theta_{\tilde{N}_{1} \tilde{\nu}_{R}}$ or $\theta_{\tilde{N}_{2} \tilde{\nu}_{I}}$ can go to zero. Small
 mixing angles, i.e., $\theta_{\tilde{N}_{1,2} \tilde{\nu}_{R,I}}\simeq \mathcal{O}(10^{-8})$ can be expected for certain parameter space. Such small mixing angles induce decay widths for $\tilde{N}_2$  and $\tilde{\chi}^\pm_1$ as low as $\mathcal{O}(10^{-15})$ GeV or less, that lead to displaced charged tracks. These tracks can be of length cm to meter that can be observed at inner tracker and calorimeters at CMS and ATLAS detectors of the LHC.

 \begin{figure}
\begin{center}
\includegraphics[width=.45\linewidth, angle=-90]{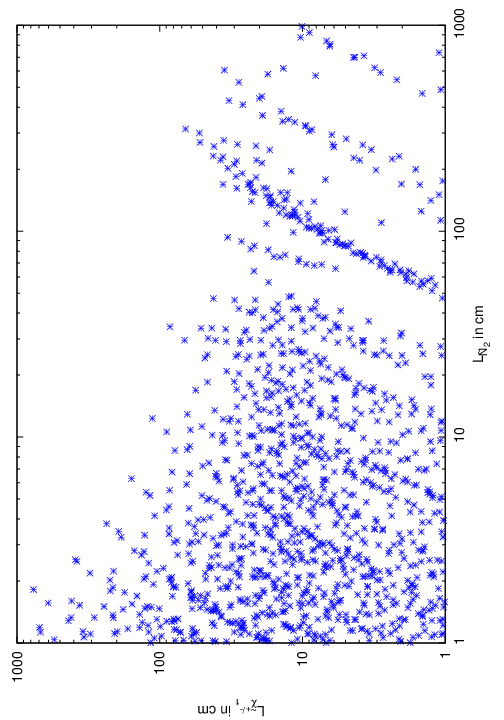}
\caption{The displaced decay lengths of $\tilde{N}_2$  and $\tilde{\chi}^\pm_1$ for values of the model parameters that can be probed at the LHC.}\label{displ}
\end{center}
\end{figure}

Figure~\ref{displ} shows such decay lengths for $\tilde{N}_2$  and $\tilde{\chi}^\pm_1$ in cm for a scanned parameter space given below in Eq.~\ref{scan}
\bea\label{scan}
100\leq m_N\leq 200 \,\rm{GeV},\quad 100\leq m_{\tilde{N}_{1,2}}\leq 500\, \rm{GeV},\quad 205\leq m_{\tilde{\nu}}\leq 505\, \rm{GeV},\nn\\
m_{\tilde{\chi}^\pm_1}\geq m_{\tilde{N}_1}+10 \, \rm{GeV},\quad V_{i1}=1, \quad \widetilde{m}^\nu=0.05\, \rm{eV}.
\eea
The parameter space is scanned for the displaced decays of lighter chargino  and NLSP  where the right-handed neutrino mass $m_N$, 
the right-handed  sneutrino masses $m_{\tilde{N}_{1,2}}$ and the left-handed sneutrino mass $m_{\tilde{\nu}}$, the lighter chargino $m_{\tilde{\chi}^\pm_1}$ and $m_{\tilde{N}_1}$ neutralino masses  are varied according to Eq.~\ref{scan} for $\widetilde{m}^\nu=0.05\, \rm{eV}$. The scan is not performed with the random variations of the parameters within the limit given in Eq.~\ref{scan} but are varied by  $50$ GeV,  $25$ GeV and $22.25$ GeV for  the senutrinos masses($m_{\tilde{N}_{1,2}}, m_{\tilde{\nu}}$), right-handed neutrino mass ($m_N$) and $X_L$ (See Eq.~\ref{snumix}) respectively. From Figure~\ref{displ} we can see that depending on the parameter space, chargino produced from the first displaced decays of $\tilde{N}_2$ can travel few cms to meters before its own displaced decay.  The little parabolic behaviour between the two decay lengths are originating due to the choice of the parameter scan $m_{\tilde{\chi}^\pm_1}\geq m_{\tilde{N}_1}+10 \, \rm{GeV}$ (See Eq.~\ref{dcn2} and Eq.~\ref{chng1}).

  
The left-handed sneutrino and sleptons are heavy enough, such that $\tilde{\chi}^\pm_1 \to \tilde{\ell}\nu, \tilde{\nu}\ell$ decays are not  kinematically allowed. In principle the lighter chargino can go through three-body decays, but the fact that the lighter RHsN ($\tilde{N}_1$) is the LSP, such decays will be further suppressed due to the following reasons.  Given the scenario, the lighter chargino ($\tilde{\chi}^\pm_1$) is wino-type and also NLSP,  the possible three-body decays are possible via off-shell selectron ($\tilde{\ell}^*$), i.e., $\tilde{\chi}^\pm_1 \to  \nu \, \tilde{\ell}^* \to \nu (\ell^\pm/W^\pm/H^\pm) \tilde{N}_1 $, via off-shell $\tilde{Z}$ or higgsino i.e., $\tilde{\chi}^\pm_1 \to  \nu \, (\tilde{Z}^*/\tilde{h}^*) \to \nu W^\pm \tilde{N}_1 $, via charged higgsino ($\tilde{H}^\pm$) i.e., $\tilde{\chi}^\pm_1 \to  h \, (\tilde{H}^{\pm *}) \to h e^\pm \tilde{N}_1 $ and via off-shell left-handed sneutrino ($\tilde{\nu}_L$), i.e., $\tilde{\chi}^\pm_1 \to  \ell^\pm \, \tilde{\nu}_L^* \to  \ell^\pm h \tilde{N}_1 $.  All the cases the three-body decay widths are proportional to the square of mixing angle of  left-right handed sneutrino ($\theta_{\tilde{N}_{1} \tilde{\nu}_{R}}$ ) and again it is further suppressed by the mass of off-shell selectron ($\tilde{\ell}^*$) or left-handed sneutrino ($\tilde{\nu}_L^*$) propagators.  However for the given scenario, the three-body decays with final states involving  $W^\pm, h, H^\pm$ are not kinematically allowed as $m_{\tilde{\chi}^\pm_1} -m_{\tilde{N}_1} < m_{W^\pm, \,h, \,H^\pm}$.  Another possibility is that of four-body decays when such $h, W^\pm, H^\pm$ are off-shell and lead to $\tilde{\chi}^+_1 \to b\bar{b} \ell^\pm \tilde{N}_1$ and $\tilde{\chi}^+_1 \to \nu \tau \nu \tilde{N}_1$. Thus the wino characteristics of the chargino makes such three-body and four-body decays more suppressed than the two-body decays.

\section{$RHN$ and $RHsN$ phenomenology}\label{pheno}

Another kind of lepton-flavour violating decay happens when the RHN $N$ decays to $e^\pm W^\mp$. This RHN can be produced from the $Z' \to NN$. However, such production rate is suppressed \cite{Bandyo11} given the recent bound on $m_{Z'}\gsim 2.5$ TeV from LHC \cite{z'bound}. The  decays of $\tilde{Z'} \to N \tilde{N}_{1,2}$ could be an alternative source of such RHN. Being electro-weak in nature, pair production of $\tilde{Z'}$ is suppressed and it is encouraged to look for $\tilde Z'$ is supersymmetric (SUSY) cascade decays \cite{Bandyo14}. From Figure 13 of \cite{Bandyo11}, we can see that the right-handed down type squark mostly decays to $\tilde{Z}^{\prime}$ for the $U(1)_\chi$ model, which is a feature of $U(1)_{\chi}$ model. This prompts us to follow the decays of third generation squarks specially 
$\tilde{t}_1$ and $\tilde{b}_1$, which are expected to be lighter and can be produced  at the LHC earlier than others as a signature of the supersymmetry. In this model the third generation squarks, in particular lighter sbottom ($\tilde{b}_1$) cascade decay to $\tilde{Z'}b$ which then follow the decay topology of Figure~\ref{fig:fy}.
\begin{eqnarray}\label{cdcy}
pp&\to& \tilde{b}_1 \tilde{b}^*_1 \to 2b+ 2\tilde{Z'} \nonumber\\
&\to& 2N + 2\tilde{N}_2 \nn \\
&\to & 2N + 2e + \tilde{\chi}^\pm_1 \nn \\
&\to& 2b+2N+4e+\etmiss
\end{eqnarray}
  Eq.~\ref{cdcy}  shows that the sbottom pair production and their following decay topologies that lead to $2b+2N+4e+\etmiss $ final states, where the two pairs of  electron come from the displaced "lepton flavour violating" decays of $\tilde{N}_2$ and $\tilde{\chi}^\pm_1$ respectively. The RHNs $N$ produced from  $\tilde{Z'}$ decay are not displaced but add to the "lepton flavour violating" decays $eW$ along with $Z\nu$ and $h\nu$ permitted by the phase space. Thus final state can be constituted of $2b+6e+2W+\etmiss$, among which four of the electrons are displaced. Now the two $W$s can decay to both muons and electrons with the same ratios. Checking the final state muon and electron numbers tagged with two $b$-jets among which 4 electrons are displaced could be a clear signature for such cascade decay.  Along with $\tilde{b}_1$ contributions, $\tilde{t}_1 \to t \tilde{Z'}$ adds to the final states with more jets or leptons  from the top decay. In the case where there is no "lepton flavour violation", we can still look for displaced charged leptons ($e$, $\mu$) tagged with prompt leptons and $b$-jets can probe such decay modes.

\begin{table}
\begin{center}
\renewcommand{\arraystretch}{1.4}
\begin{tabular}{||c||c|c|c||}
\hline\hline
mass&BP1&BP2&BP3\\
\hline\hline
$m_{\tilde{b}_{1}}$& 634.3 &656.2 &622.7\\
\hline
$m_{\tilde{t}_{1}}$ & 561.5 &547.4 &543.5\\
\hline
$m_{\tilde{Z'}}$ & 300.0 &250.0 &300.0\\
\hline
$m_{\tilde{N}_{2}}$& 150.0 & 140.0 &150.0\\
\hline
$m_{\tilde{\chi}^\pm_{1}}$& 121.0 & 130.0 &140.0\\
\hline
$m_{\tilde{N}_{1}}$& 110.0 & 100.0 &110.0\\
\hline
$m_{N}$& 100.0 &100.0 &130.0\\
\hline
$X_L$& 89.9 &89.0&89.9\\
\hline\hline
\end{tabular}
\caption{\label{bps}The benchmark points for the collider
study where $\widetilde{m}^\nu=0.05$ eV. }
\end{center}
\end{table}

For the collider study we select few benchmark points given in Table~\ref{bps}
where we mainly follow Ref~\cite{Bandyo14} for fixing the Higgs sector with decoupled
first two generations of squark and gluino.   The recent bounds from LHC 
on Higgs data \cite{LHCH}, bounds on third generation SUSY masses \cite{thirdgensusy} 
and recent bounds on lighter charginos and neutralinos production cross-sections \cite{chgn1}
 are taken into account. The mass spectrum in the benchmark points generally follows the hierarchy shown in Figure~\ref{massspec}, except for BP3 where the right-handed neutrino $N$ decay to $h\nu$ is kinematically allowed unlike the other two benchmark points.

\begin{table}
\begin{center}
\renewcommand{\arraystretch}{1.4}
\begin{tabular}{||c||c|c|c||}
\hline\hline
Displaced&\multicolumn{3}{|c||}{Decay length in cm}\\
 \cline{2-4}
decays & BP1&BP2&BP3\\
\hline\hline
$\tilde{N}_2$ &795.8&736.9&50.5\\
\hline
$\tilde{\chi}^\pm_1$&404.6&69.9&53.6\\
\hline\hline
\end{tabular}
\caption{\label{dcyl}The displaced decay length of $\tilde{N}_2$ and $\tilde{\chi}^\pm_1$ for the benchmark points. }
\end{center}
\end{table}
 In this collider study we mainly focus on the multi-leptonic final states
 tagged with at least two $b$-jets ($2b+ \geq 6\ell$) along with lepton flavour, i.e. $e$ and $\mu$.  Four of the electrons coming from 
 $\tilde{N}_2$ and $\tilde{\chi}^\pm_1$  are displaced and their average decay lengths are shown
 in Table~\ref{dcyl}.   Table~\ref{br} shows the relevant decay branching fractions involved in SUSY cascade decays.  For BP1 and BP2 the RHN is 100 GeV so decays to $W^\pm e^\mp$ and $Z\nu$ with branching fraction of 88\%, 12\% respectively. However, for BP3 as $m_N=130$ GeV it decays to $h\nu$ as well with branching fraction of $\sim 0.3\%$. 
 
\begin{table}
\begin{center}
\renewcommand{\arraystretch}{1.4}
\begin{tabular}{||c|c|c|c||}
\hline\hline
mass&BP1&BP2&BP3\\
\hline\hline
$\mathcal{B}(\tilde{b}_{1}\to b\tilde{Z'})$& 1.00 &0.11 &1.00\\
\hline
$\mathcal{B}(\tilde{t}_{1}\to b\tilde{Z'})$ & 1.00 &0.02 &1.00\\
\hline
$\mathcal{B}(\tilde{Z'}\to N\tilde{N}_2)$ & 0.40&0.28 &0.35\\
\hline
$\mathcal{B}(\tilde{Z'}\to N\tilde{N}_1)$ & 0.60 &0.72&0.65\\
\hline\hline
\end{tabular}
\caption{\label{br} Relevant Branching fractions in supersymmetric cascade decay
for  the benchmark points. }
\end{center}
\end{table}

 \section{Collider simulation}\label{cold}
  
 For this purpose we simulate the events coming from stops and sbottoms via CalcHEP-PYTHIA\cite{calchep,pythia}  interface for the final states. The jet formation has been performed using the {\tt Fastjet-3.0.3} \cite{fastjet} with the {\tt CAMBRIDGE AACHEN} algorithm. We have selected a jet size $R=0.5$ for the jet formation, with the following criteria:
\begin{itemize}
  \item the calorimeter coverage is $\rm |\eta| < 4.5$

  \item the minimum transverse momentum of the jet $ p_{T,min}^{jet} = 20$ GeV and jets are ordered in $p_{T}$
  \item leptons ($\rm \ell=e,~\mu$) are selected with
        $p_T \ge 20$ GeV and $\rm |\eta| \le 2.5$
  \item no jet should be accompanied by a hard lepton in the event
   \item $\Delta R_{lj}\geq 0.4$ and $\Delta R_{ll}\geq 0.2$
  \item Since an efficient identification of the leptons is crucial for our study, we additionally require  
a hadronic activity within a cone of $\Delta R = 0.3$ between two isolated leptons to be $\leq 0.15\, p^{\ell}_T$ GeV, with 
$p^{\ell}_T$ the transverse momentum of the lepton, in the specified cone.

\end{itemize}

\begin{figure}
\begin{center}
\mbox{\subfigure[]{
\includegraphics[width=.33\linewidth, angle=-90]{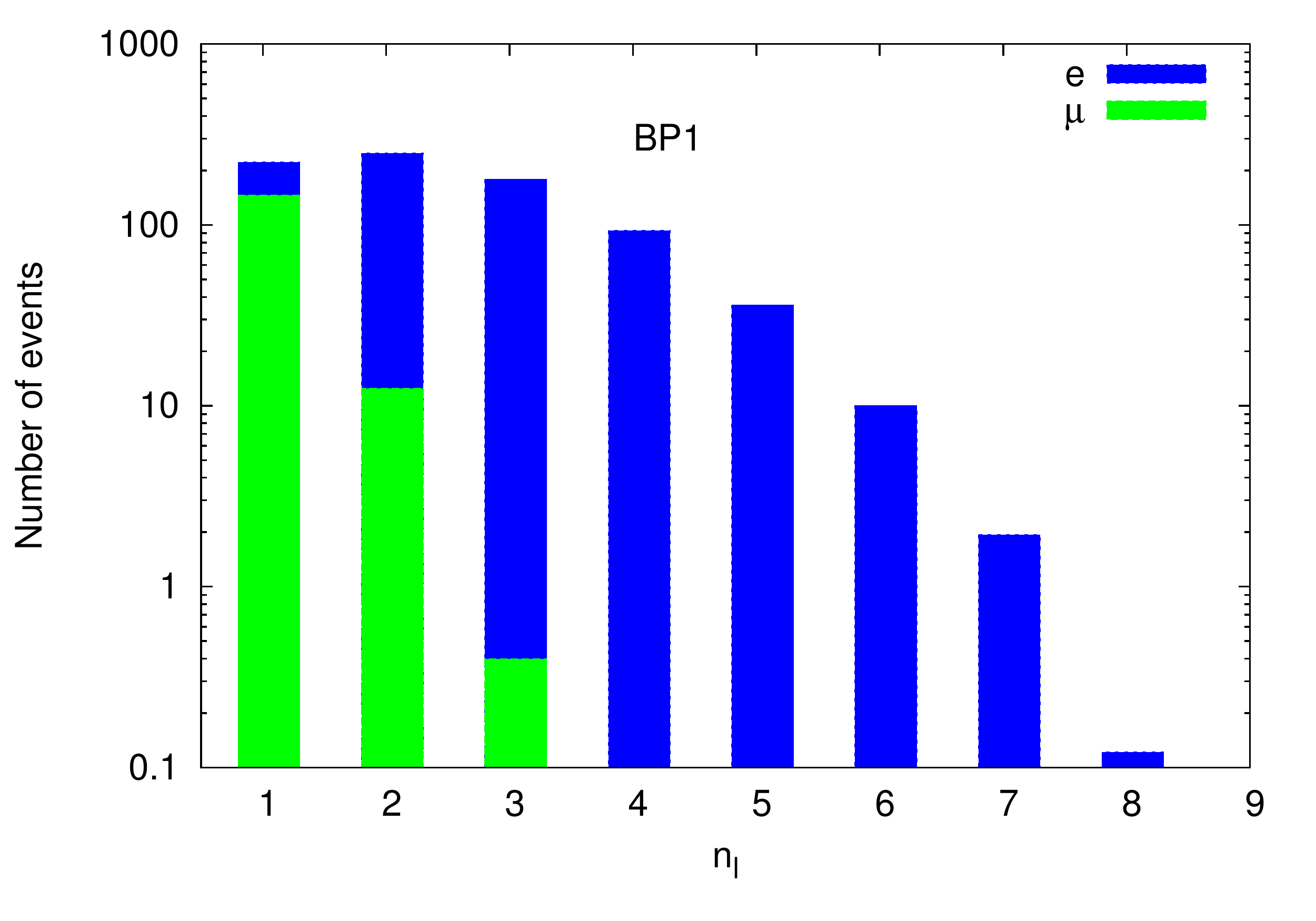}}
\subfigure[]{
\includegraphics[width=.33\linewidth, angle=-90]{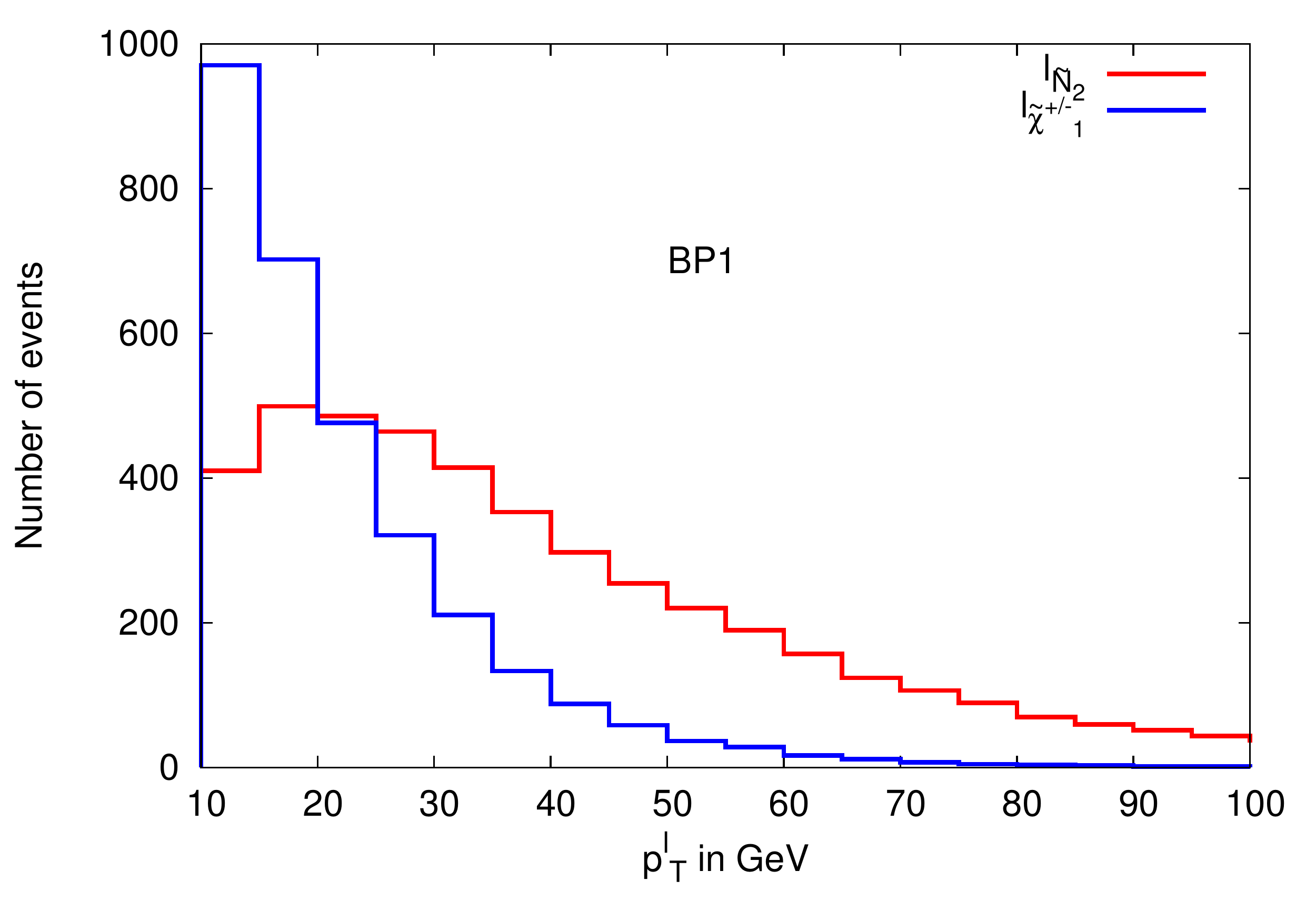}}}
\caption{Electron and muon multiplicity distribution (a) and their $p_T$ distribution (b) at the LHC with 14 TeV.}\label{lmlbt}
\end{center}
\end{figure}

\begin{figure}
\begin{center}
\includegraphics[width=.45\linewidth, angle=-90]{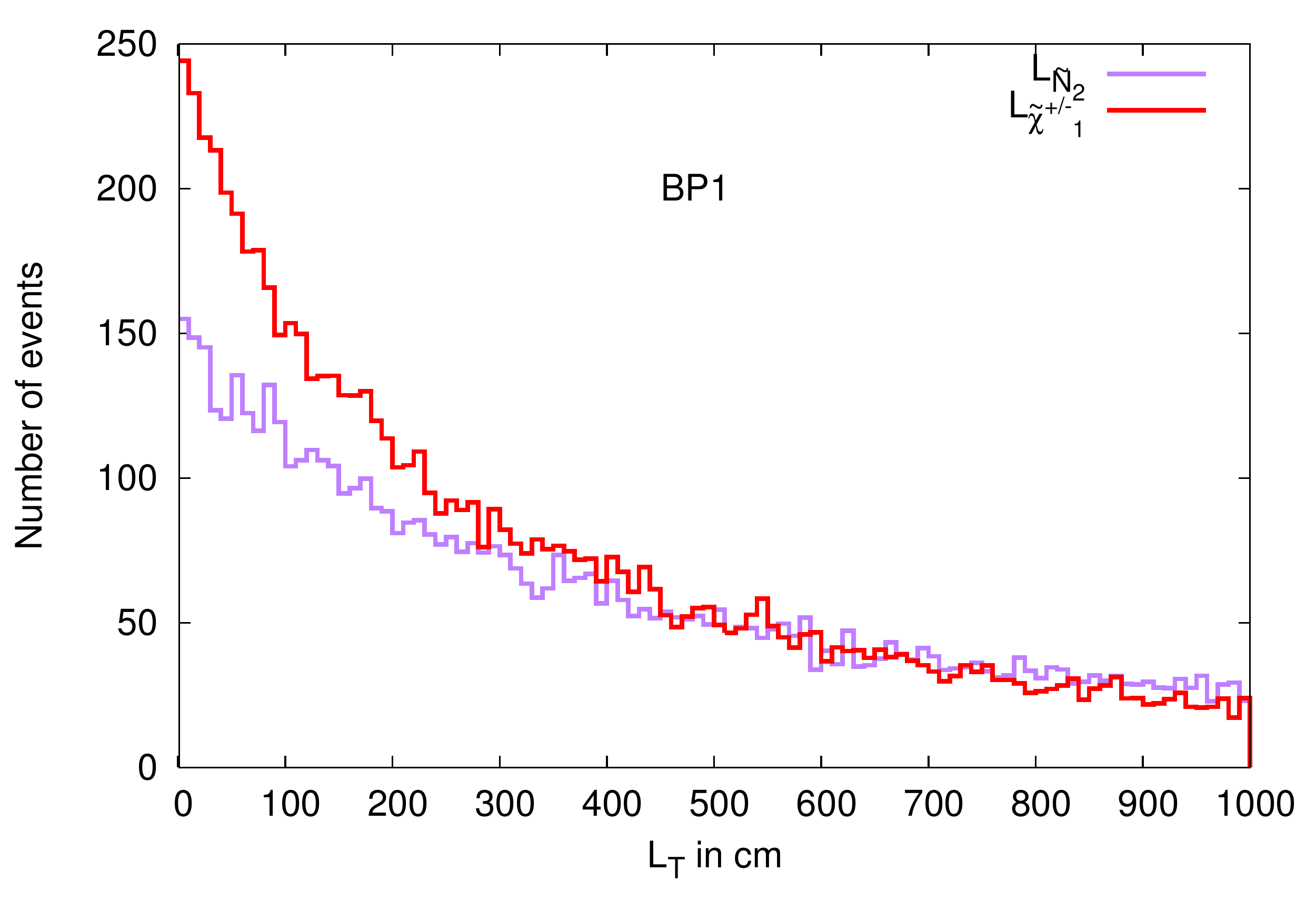}
\caption{Displaced decay length of $\tilde{N}_2$ and $\tilde{\chi}^\pm_1$ at the LHC with 14 TeV.}\label{disim}
\end{center}
\end{figure}

Figure~\ref{lmlbt} (a) shows the electron and muon number multiplicity distribution from $\tilde{b}_1 \tilde{b}^*_1$ for BP1 and it is evident that due to "lepton flavour violating" decays, electron numbers are more than muons. Figure~\ref{lmlbt} (b) shows the $p_T$ distributions of such electrons coming from 
$\tilde{N}_2$ and $\tilde{\chi}^\pm_1$ and they could be on softer side due to smaller mass gaps in the corresponding "lepton flavour violating" decays. Figure~\ref{disim} shows the transverse decay lengths of 
$\tilde{N}_2$ and $\tilde{\chi}^\pm_1$ for BP1 at the LHC for ECM of 14 TeV.

\begin{table}
\begin{center}
\renewcommand{\arraystretch}{1.4}
\begin{tabular}{||c|c|c|c|c||}
\hline\hline
Final states&Production &BP1&BP2&BP3\\
 &modes&&&\\
\hline
\multirow{2}{*}{$\geq 4\ell +2b$-jets}&$\tilde{b}_1\tilde{b}^*_1$&1173.8 &74.8& 699.4\\
& $\tilde{t}_1\tilde{t}^*_1$&1310.2&19.0& 1327.7\\
\hline
\multirow{2}{*}{$\geq 4e +2b$-jets}&$\tilde{b}_1\tilde{b}^*_1$&910.1 &53.0& 531.0\\
& $\tilde{t}_1\tilde{t}^*_1$&867.8&11.1& 821.7\\
\hline
\multirow{2}{*}{$\geq 4\mu +2b$-jets}&$\tilde{b}_1\tilde{b}^*_1$&0.0&0.0& 0.0\\
& $\tilde{t}_1\tilde{t}^*_1$&0.2&0.0& 2.6\\
\hline
\hline
\end{tabular}
\caption{\label{fnst144l} Final state numbers for $4\ell, 4e, 4\mu$ along with $2b$-jets
for the benchmark points at 100 fb$^{-1}$at the LHC with 14 TeV, where the contributions are from $\tilde{b}_1\tilde{b}^*_1$ and $\tilde{t}_1\tilde{t}^*_1$ respectively.}
\end{center}
\end{table}

\begin{table}
\begin{center}
\renewcommand{\arraystretch}{1.4}
\begin{tabular}{||c||c||}
\hline\hline
Final &$tTZ$\\

states&at 14 TeV\\
\hline
$\geq 4\ell +2b$-jets&20.8\\
\hline
$\geq 4e +2b$-jets &2.6\\
\hline
$\geq 4\mu +2b$-jets&1.0\\
\hline
\hline
\end{tabular}
\caption{\label{fnst144bg} Final state numbers for $4\ell, 4e, 4\mu$ along with $2b$-jets
for the dominant SM background $tTZ$ at 100 fb$^{-1}$ at the LHC with 14 TeV, where all the charged leptons are prompt ones.}
\end{center}
\end{table}

Table~\ref{fnst144l} presents the number of events for the benchmark 
points for the final states $\geq 4\ell +2b$-jets, $\geq 4e +2b$-jets and $\geq 4\mu +2b$-jets at an integrated luminosity of 100 fb$^{-1}$ for the center of mass energy of 14 TeV at the LHC. Here $\ell$ includes $e$ and $\mu$
and we take the single $b$-jet tagged efficiency of 0.5 \cite{btag}. We can see that
that $\geq 4e+ 2b$-jets has greater numbers of events compared to $\geq  4\mu +2b$-jets. This is due to 
the "lepton flavour violating" decays of  $\tilde{N}_2$  and $\tilde{\chi}^\pm_1$, from where four electrons are  coming from. Whereas the muons are coming from the $W^\pm$s which are produced from the decays of RHNs ($N$). The difference between electron and muon numbers obviously gives the "lepton flavour violating" signature.  On top of that, four of the electrons from $\tilde{N}_2$  and $\tilde{\chi}^\pm_1$ decays are displaced ones.  If the lepton flavour is conserved then we expect same number of displaced electrons and muons in the final sates. The existence of displaced charged leptonic final states 
make the final state topologies almost background free. In terms of charged lepton numbers and $b$-jets,
there can be some contributions for $4\ell +2b$-jet final state from SM $ZZZ$, $t\bar{t}Z$ but without any displaced charged leptons. However, $ZZZ$ background fails to contribute after the selection cuts and due to low cross-section.  Table~\ref{fnst144bg} presents the corresponding $t\bar{t}Z$ number for  14 TeV at the LHC, where all the charged leptons are prompt ones. The signals remain clean if we consider the displaced charged leptons due to no SM backgrounds.  However, considering all the charged leptons are prompt we can calculate the signal significance for the benchmark points where the cascade decay branching fractions affect the
final state numbers. For $4\ell +2b$-jet final state final state gets 
signal significance of 49.6$\sigma$, 8.6$\sigma$, and 44.7$\sigma$ respectively for center of mass energy of 14 TeV at the LHC. A $5\sigma$ discovery can be achieved with very earlier data of few fb$^{-1}$ for BP1 and BP3 to 34 fb$^{-1}$ for BP2 for 14 TeV at the LHC.  $4e +2b$-jet final state get significance of 42.1$\sigma$,  7.8$\sigma$ and  36.7$\sigma$ respectively for BP1, BP2 and BP3 at the LHC with ECM=14 TeV. Due to "lepton flavour violating" decays $(4e -4\mu) +2b$-jet final state will have greater significance than $4e+2b$-jet case, which can be realised from Table~\ref{fnst144l} and Table~\ref{fnst144bg}.

Table~\ref{fnst146l} presents $\geq 6\ell$ final states event numbers for the benchmark points at the LHC with ECM=14 TeV at an integrated luminosity of 100 fb$^{-1}$, which also shows $\geq 6e$ and $\geq 4e + \geq 2\mu$ final states. The final states are almost background free specially with the displaced charged electrons. For $\geq 6\ell $ final state final state gets 
signal significance of 29.4$\sigma$, 4.2$\sigma$, and 25.1$\sigma$ respectively for BP1, BP2 and BP3 for center of mass energy of 14 TeV at the LHC at an integrated luminosity of 100 fb$^{-1}$. The required luminosity for $5\sigma$ significance can vary from few fb$^{-1}$ to $\mathcal{O}(200)$ fb$^{-1}$. $\geq 6\ell +2b$-jets final state signal significance numbers are 12.8$\sigma$, 2.1$\sigma$, and 10.1$\sigma$ respectively for BP1, BP2 and BP3 for center of mass energy of 14 TeV at the LHC at an integrated luminosity of 100 fb$^{-1}$. Similarly, we can see that 
for $\geq 4\ell +\geq 2\mu +2b$-jets the corresponding significances are 1.0$\sigma$, 1.7$\sigma$, and 10.1$\sigma$ respectively for BP1, BP2 and BP3 for center of mass energy of 14 TeV at the LHC at an integrated luminosity of 100 fb$^{-1}$.

\begin{table}
\begin{center}
\renewcommand{\arraystretch}{1.4}
\begin{tabular}{||c|c|c|c|c||}
\hline\hline
Final states&Production &BP1&BP2&BP3\\
 &modes&&&\\
\hline
\hline
\multirow{2}{*}{$\geq 6\ell $}& $\tilde{b}_1\tilde{b}^*_1$&215.3&10.7&102.0\\
& $\tilde{t}_1\tilde{t}^*_1$&646.6&6.9& 527.1\\
\hline
\multirow{2}{*}{$+2b$-jets}&$\tilde{b}_1\tilde{b}^*_1$ &137.8&7.0& 66.0\\
& $\tilde{t}_1\tilde{t}^*_1$ &194.1&2.2&168.2\\
\hline
\multirow{2}{*}{$\geq 6e+2b$-jets}&$\tilde{b}_1\tilde{b}^*_1$ &79.3 &3.5& 35.4\\
& $\tilde{t}_1\tilde{t}^*_1$ &84.8&0.8&66.1\\
\hline
$\geq 4e+\geq 2\mu$&$\tilde{b}_1\tilde{b}^*_1$&32.6&1.7&14.6\\
$+2b$-jets& $\tilde{t}_1\tilde{t}^*_1$&86.8&1.2&87.8\\
\hline\hline
\end{tabular}
\caption{\label{fnst146l} Final state numbers for  the benchmark points at 100 fb$^{-1}$
at the LHC with 14 TeV, where the contributions are from $\tilde{b}_1\tilde{b}^*_1$ and $\tilde{t}_1\tilde{t}^*_1$ respectively.}
\end{center}
\end{table}

 BP1 and BP3 cases are favourable points to probe the full decay chain  including the electrons coming from the RHNs and electrons and muons from the $W^\pm$ bosons along with the four displaced electrons from chargino and RHsN decays. When both the $W^\pm$s decay leptonically we get final states with $8e$ and $6e+2\mu$. Finally in Table~\ref{fnst148l} we present the number of events for the benchmark 
points for the final states at an integrated luminosity of 100 fb$^{-1}$ for the center of mass energy of 14 TeV at the LHC respectively. Though for higher charged leptonic final states signal numbers suffer a lot but finding such multi-leptonic final state is still a possibility which can probe this particular decay chain.  
If such a decay chain with displaced charged lepton can be found then we can measure the mixings between 
the lighter RHsNs  with the left-handed at the LHC.

\begin{table}
\begin{center}
\renewcommand{\arraystretch}{1.4}
\begin{tabular}{||c|c|c|c|c||}
\hline\hline
Final states&Production &BP1&BP2&BP3\\
 &modes&&&\\
\hline
\hline
\multirow{2}{*}{$\geq 8e$}&$\tilde{b}_1\tilde{b}^*_1$&1.2&0.1&1.0\\
& $\tilde{t}_1\tilde{t}^*_1$ &7.9&$10^{-2}$&7.0\\
\multirow{2}{*}{$+2b$-jets}&$\tilde{b}_1\tilde{b}^*_1$&0.7&0.1&0.6\\
& $\tilde{t}_1\tilde{t}^*_1$ &1.5&$10^{-3}$&3.7\\
\hline
\multirow{2}{*}{$\geq 6 e+\geq 2\mu$}&$\tilde{b}_1\tilde{b}^*_1$&1.8&0.1&0.6\\
& $\tilde{t}_1\tilde{t}^*_1$&12.0&0.1&10.3\\
\multirow{2}{*}{$+2b$-jets}&$\tilde{b}_1\tilde{b}^*_1$&1.2&0.1&0.5\\
& $\tilde{t}_1\tilde{t}^*_1$ &3.1&$10^{-2}$&2.6\\
\hline\hline
\end{tabular}
\caption{\label{fnst148l} Final state numbers for  the benchmark points at 100 fb$^{-1}$
at the LHC with 14 TeV, where the contributions are from $\tilde{b}_1\tilde{b}^*_1$ and $\tilde{t}_1\tilde{t}^*_1$ respectively.}
\end{center}
\end{table}

\section{Discussion and conclusion}\label{concl}
We have seen in this study that displaced leptonic signatures can probe the left-right handed sneutrino 
mixings as well as the possible "lepton flavour violation" at the LHC.  Here we analysed multi-leptonic final state
with some displaced charged leptonic signature. The choice of parameter space is responsible for such displaced 
decays. However, these multi-leptonic final states can address other parameter space without having a dispalced decay. In a $\tilde{Z}'$ NLSP scenario,  $\tilde{Z}' \to N\tilde{N}_1\to e^\pm W^\mp \tilde{N}_1$ decay produces more prompt electron than muon in the final states and such "lepton flavour violation" if any can be measured by checking the flavour multiplicities in the final states \cite{Bandyo14} which can also give rise to needed same sign di-leptonic excess recently found at the LHC \cite{SSD}. The nature of the compact mass spectrum as well as the nature of NLSP changes the scenario quite a bit which we discuss below.

In the present case compact mass spectrum is responsible for the final state resulting two displaced charged leptons from the decays of $\tilde{Z'}$ NNLSP and of wino-like chargino NLSP. The fact is that $\tilde{N}_{1,2}$ only mix via $\tilde{\nu}_L$ (see Eq.~\ref{snumix}) and couple to wino-like chargino. As long as $\tilde{Z'}$ is NNLSP, wino-like chargino is NLSP  and $\tilde{N}_1$ is LSP, the above signature is normally expected.  Even if the mass gap between chargino and LSP is greater than $m_W$, $\tilde{W}^\pm \to W^\pm \tilde{N}_1$ is not allowed due to the absence of direct coupling. However in such situation, the three-body displaced decays, i.e., $\tilde{W}^\pm \to W^\pm \nu (\ell) \tilde{N}_1$ open up via off-shell charged slepton, $\tilde{Z}$ and higgsino respectively. The displaced $W^\pm$ then decays promptly into charged leptons and neutrinos or di-jet. For an increased mass gap between $\tilde{W}^\pm$ and $\tilde{N}_1$, i.e., $(m_{\tilde{W}^\pm} -m_{\tilde{N}_1}) > m_h$, a displaced Higgs production may be visible via the three-body decay of $\tilde{W}^\pm \to \ell^\pm h \tilde{N}_1$. Similar argument also holds for the displaced decay $\tilde{N}_2\to \ell \tilde{W}^\mp$. As long as the NLSP is wino-like chargino, no other kind of two-body decay is allowed even if the mass gap is more than $m_W$. However if the mass gap is more than the Higgs($h$) mass, $\tilde{N}_2 \to h \,\ell^\mp\, \tilde{W}^\mp$ is open due to the existence of $\tilde{N}_{1,2} - h- \tilde{\nu}$ couplings, which leads to displaced Higgs production along with a displaced leptonic charged track. 

If we allow charged slepton ($\tilde{\ell})$ instead of wino-like chargino as NLSP, then also we can have displaced charged lepton or $W^\pm$ in one or both the vertices depending on the mass spectrum.  For a $\tilde{Z}$ like NLSP  scenario, such displaced decays of both $\tilde{N}_2$ and $\tilde{Z}$ will give rise to neutrinos and $\tilde{N}_1$, i.e. only the missing particles and thus not recognisable at the collider. The signature of displaced charged lepton or $W^\pm$ changes drastically if $\tilde{Z'}$ is NLSP and decays to $N\tilde{N}_1$, which gives rise to non-displaced "lepton flavour violating" signature at the collider \cite{Bandyo14}.  If the lighter chargino is $\tilde{H}^\pm_u$ type, then $\tilde{N}_2 \to l\tilde{H}^\pm_u$ decay width is proportional to the $y^2_{\nu}$ which may also lead to displaced decay of $\tilde{N}_2$. However, given $\tilde{N}_1$ is LSP, such higgsino like chargino will decay to $\ell \, \tilde{N}_1$ via $\theta_{\tilde{N}_{1} \tilde{\nu}_{R}}$  and travel a few cms to meters before encountering a recoil in it's charged track. 

These final states are still interesting even when $\tilde{N}_2$ and $\tilde{\chi}^\pm_1$ do not have "lepton flavour violating" final states, i.e. they decay into the electron, muon and tau channels equally.  Thus displaced electron and muon number difference is a good observable to probe such scenarios. In case of supersymmetric Type III seesaw, a pair of charged RHsN have similar interesting phenomenology\cite{Bandyo10}.  There are some other studies where right-handed sneutrino is considered as cold dark matter  \cite{rhsndm} and similar long lived charged particles are predicted from supersymmetric cascade decays \cite{lit}.

The displaced charged leptonic signature can appear in various supersymmetric models,  viz. $R$-parity violation \cite{rpvd}, gauge mediation \cite{gmd}, etc  and non-supersymmetric models, viz. Type-I seesaw \cite{ty1d}, Hyperpions \cite{hypd}, Leptoquark \cite{lqd}, etc. Thus there is a wide range of models that can mimic the generic signature of displaced decays and the corresponding charged tracks. However, we see in this article that the presence of one of right-handed sneutrino as dark matter makes the production of the corresponding right-handed neutrino dominant over other two right-handed neutrino flavours. This brings not only displaced charged tracks, but also lepton-flavour violating signatures. In that respect this is a "smoking gun" signature which can distinguish such model from others with generic displaced leptonic charged track.

\begin{acknowledgments}

PB wants to acknowledge  Eung Jin Chun for the useful discussions.
\end{acknowledgments}


\end{document}